\documentclass{webofc}
\usepackage[varg]{txfonts}   
\usepackage{graphicx,amsmath,amssymb,float,hyperref}
 
\begin{document}

\title{Transverse momentum fluctuation in ultra-central Pb+Pb collision at the LHC}

\author{\firstname{Rupam} \lastname{Samanta}\inst{1,2}\fnsep\thanks{\email{rsamanta@agh.edu.pl}} \and
        \firstname{Somadutta} \lastname{Bhatta}\inst{3} \and
        \firstname{Jiangyong} \lastname{Jia}\inst{3,4} \and
        \firstname{Matthew} \lastname{Luzum}\inst{5} \and
        \firstname{Jean-Yves} \lastname{Ollitrault}\inst{2}\fnsep
}

\institute{AGH University of Science and Technology, Faculty of Physics and Applied Computer Science, aleja Mickiewicza 30, 30-059 Cracow, Poland
\and
           Universit\'e Paris Saclay, CNRS, CEA, Institut de physique th\'eorique, 91191 Gif-sur-Yvette, France
\and
           Department of Chemistry, Stony Brook University, Stony Brook, NY 11794, USA
\and
           Physics Department, Brookhaven National Laboratory, Upton, NY 11976, USA
\and
           Instituto de F\'{\i}sica, Universidade de  S\~{a}o Paulo,  Rua  do  Mat\~{a}o, 1371,  Butant\~{a},  05508-090,  S\~{a}o  Paulo,  Brazil
          }

\abstract{
The ATLAS collaboration has recently observed that the variance of the transverse momentum per particle ($[ p_t ]$), when measured as a function of the collision multiplicity ($N_{ch}$) in Pb+Pb collisions, decreases by a factor $2$ for the largest values of $N_{ch}$, corresponding to ultra-central collisions.  
We show that this phenomenon is naturally explained by invoking impact parameter ($b$) fluctuations, which contribute to the variance, and gradually disappear in ultra-central collisions.  
It implies that $N_{ch}$ and $[ p_t ]$ are strongly correlated at fixed $b$, which is explained by the local thermalization of the QGP medium.}

\maketitle

\section{Introduction}
\label{data}
\begin{figure}[h]
\centering
\includegraphics[height = 5 cm]{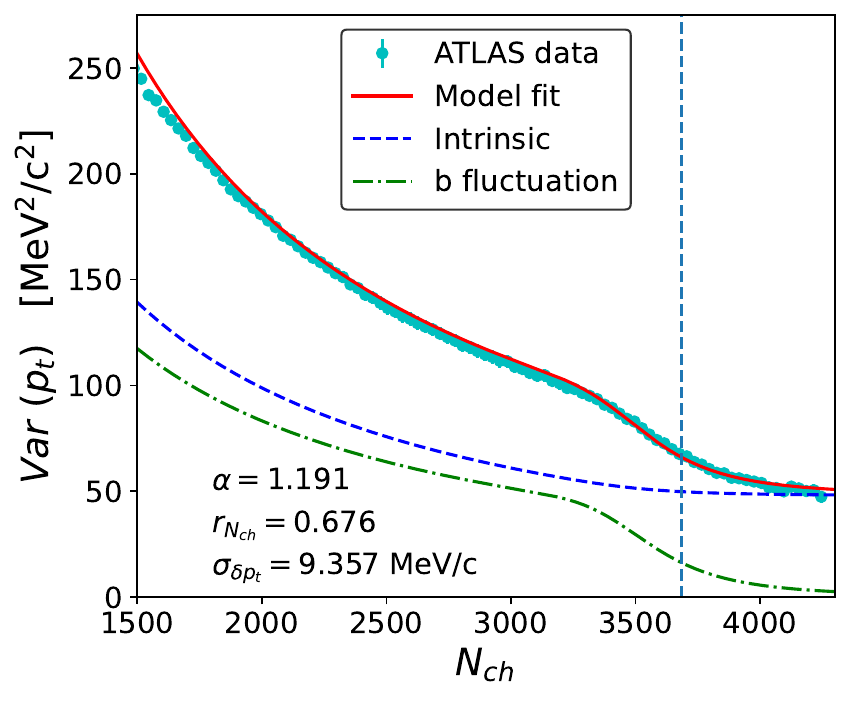}~~~~~
\includegraphics[height = 5 cm]{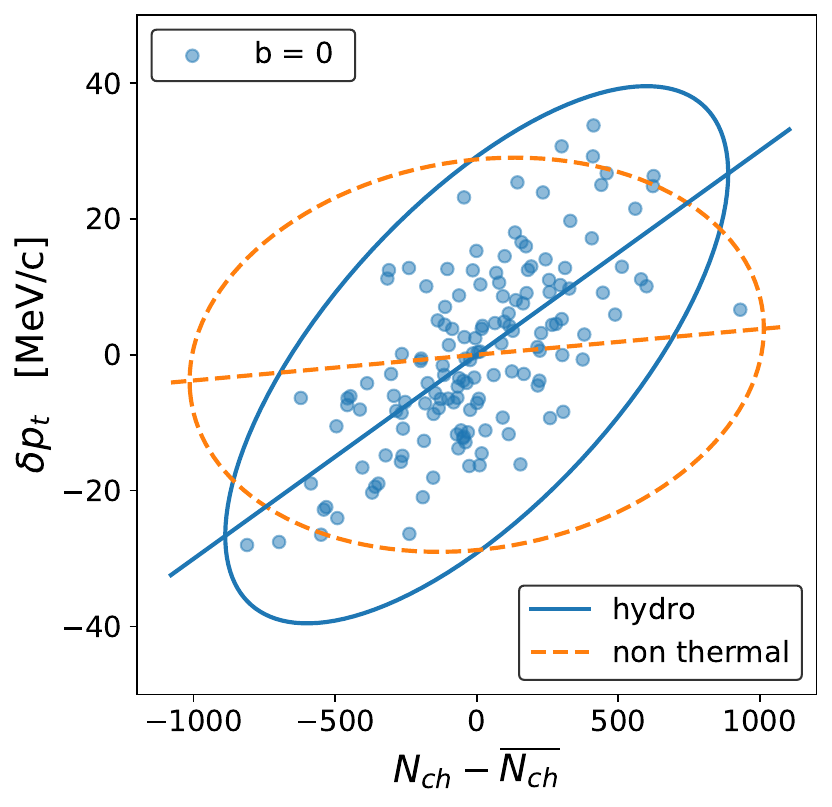}
\caption{Left: ATLAS data (light blue symbols) and our model fit presented in Sec.\ref{fit to data} (solid red curve). 
The green curve denotes the contribution due to $b$ fluctuation and the blue dashed line denotes the contribution of intrinsic fluctuation. 
The vertical dashed line is the knee of the multiplicity distribution. 
Right : Two dimensional distribution of $[p_t]$ and $N_{ch}$ from simulation of Pb+Pb collisions at 5.02 TeV and b=0. 
Blue symbols are hydrodynamic calculations. 
Instead of $[p_t]$ and $N_{ch}$, we plot the differences $N_{ch}-\overline{N_{ch}}$ and $\delta p_t = [p_t]-\overline{p_t}$ where $\overline{N_{ch}}=6692$ and $\overline{p_t}=1074 MeV/c$ are the average values. 
The blue curve is the 99\% confidence ellipse of a Gaussian distribution fitted to these symbols. 
The orange curve is a similar 99\% confidence ellipse for 1.4 $\times$ 10$^6$ collisions simulated with HIJING.  
The straight lines represent the average values $\delta p_t (N_{ch},b=0)$.}
 \vskip -.5truecm
\label{fig-1}      
\end{figure}

The left panel of Fig. \ref{fig-1} displays the variance of transverse momentum per particle ($[p_t]$) as a function of the collision multiplicity in Pb+Pb collisions at the LHC \cite{ATLAS:2022dov}. 
(It is the dynamical contribution to the variance, after subtracting the trivial statistical fluctuation from the finite multiplicity.)
It decreases over a narrow range of multiplicity around $N_{ch} = 3700$. 
This phenomenon is not explained by models such as HIJING, which predicts a smooth $1/N_{ch}$ dependence of the variance for all ranges of $N_{ch}$~\cite{Bhatta:2021qfk}. 
We show that it is naturally explained if the medium produced during the collison thermalizes. 

\section{Hydro vs HIJING results}
\label{simulation}
The impact parameter of the collision, $b$, is not measured, and varies event by event. 
Fluctuations observed in a sample of collisions event originate from fluctuations of $b$, and from fluctuations at fixed $b$. 
In order to understand their respective roles, we run simulations of  Pb+Pb collisions at 5.02 TeV at fixed $b$, more specifically at $b=0$. 

First, we simulate 150 events using relativistic viscous hydrodynamics \cite{Bozek:2021mov} and compute $N_{ch}$ and $[p_t]$ in each event. 
Fig. \ref{fig-1}  displays the distribution of the two quantities from hydrodynamic simulation where thermalization of the QGP is assumed. 
It is clearly visible that both $[p_t]$ and $N_{ch}$ have significant fluctuation. 
As the impact parameter is fixed, the origin of these fluctuations is purely quantum, due to genuine intrinsic fluctuation at the initial state of the collision. 

Most importantly, there is a strong positive correlation between $[p_t]$ and $N_{ch}$, whose physical origin is clear: 
Larger $N_{ch}$ means larger density $N_{ch}/V$, as the collision volume, which is mostly defined by the impact parameter of the collision, is fixed. 
A larger density results in a larger initial temperature if the system thermalizes. 
This in turn implies a higher energy per particle in the final state, and eventually a larger transverse momentum per particle $[p_t]$, generating the positive correlation between $[p_t]$ and $N_{ch}$.

Second, we run simulations using the HIJING model~\cite{Gyulassy:1994ew}, which considers the collision as a superposition of independent nucleon-nucleon collision, and in which the particles have no interaction between each other after production. 
The striking difference in comparison to the hydro results is that the correlation between $N_{ch}$ and $[p_t]$ is smaller by a factor $\sim$ 10.   
In HIJING, a larger density does not imply a significant increase of $[p_t]$, which can be ascribed to the lack of thermalization. 

\section{Model fit}
\label{fit to data}
\begin{figure}[h]
	\centering
	\includegraphics[height = 4.5 cm]{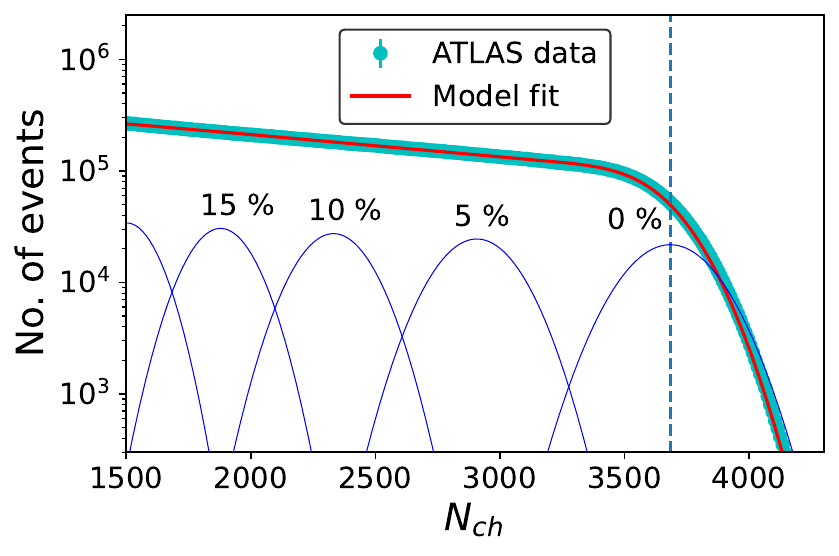}~
        \includegraphics[height = 4.5 cm]{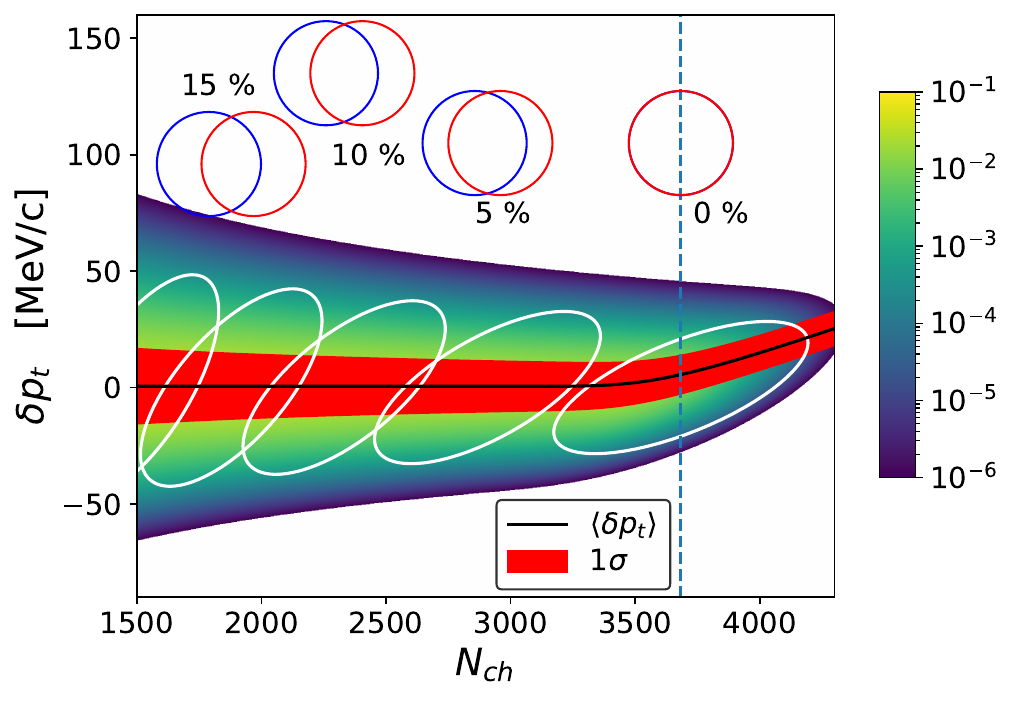}
	\caption{Left : Charged particle multiplicity ($N_{ch}$) distribution in Pb+Pb collision at 5.02 TeV measured by the ATLAS collaboration, denoted by the light blue symbols. The solid red line represents the model fit. The thin blue lines corresponds to the Gaussian distributions $P(N_{ch} | b)$ drawn for 0 \%, 5 \%, 10 \% and 15 \% centrality fractions. The vertical dashed line corresponds to the "knee" ($\overline{N_{ch}{b=0}}$) of the distribution. Right: Two dimensional joint distribution of transverse momentum per particle $[p_t]$ and multiplicity $N_{ch}$ returned by the model fit to the fluctuation data. Rather than plotting $[p_t]$, we plot $\delta p_t=[p_t] - \overline{p_t}(0)$. The white curves represent the 99 \% confidence ellipses at fixed impact parameters. Corresponding to these values $b$, a schematic representation of the colliding nuclei is shown. The solid black line denotes the mean value of $\delta p_t$ averaged over events and the red band represents 1-$\sigma$ band.}
	\label{fig-2}       
 \vskip -.5truecm
\end{figure}

We model fluctuations of $[p_t]$ by assuming that the joint probability distribution of $[p_t]$ and $N_{ch}$ at fixed $b$ is a two-dimensional correlated Gaussian~\cite{Samanta:2023amp}, denoted by $P([p_t],N_{ch}| b)$. 
It is characterized by five parameters: mean and variance of $[p_t]$ and $N_{ch}$, denoted by $\overline{p_t}(b)$, $\overline{N_{ch}}(b)$, Var($p_t | b$), Var($N_{ch}|b$) and the Pearson correlation coefficient between $[p_t]$ and $N_{ch}$, $r_{N_{ch}}(b)$ which we expect to be positive in hydrodynamics, as shown in Fig. \ref{fig-1} (right). 

The mean and variance of $N_{ch}$ are obtained by fitting the measured distribution of $N_{ch}$ as a superposition of Gaussians, along the lines of Ref.~\cite{Das:2017ned}. 
The average value of $N_{ch}$ at $b=0$ is the knee of the distribution, which is reconstructed precisely. 
Note that the fall of the variance in Fig. \ref{fig-1} precisely occurs around the knee. 

The mean transverse momentum $\overline{p_t}(b)$ is essentially independent of $b$ for 30 \% most central collisions \cite{ALICE:2018hza}. 
We subtract this constant and only study the distribution of the distribution of $\delta p_t$ = $[p_t]-\overline{p_t}$. 
We assume that the variance Var($p_t | b$) varies with the mean according to a power law, Var($p_t | b$) = $\sigma^2_{\delta p_t} (\overline{N_{ch}}(0)/\overline{N_{ch}}(b) )^\alpha$, where $\alpha$ and $\sigma_{\delta p_t}$ are constants \cite{Samanta:2023amp}. 
For simplicity, we assume the correlation $r_{N_{ch}}$ to be independent of $b$.

The variance of $\delta p_t$ at fixed $N_{ch}$ is then evaluated in the following way. 
We first obtain the distribution of $\delta p_t$ at fixed $N_{ch}$ {\it and\/} $b$ using $P(\delta p_t | N_{ch}, b) = P(\delta p_t , N_{ch} | b) / P( N_{ch} | b)$. 
It is again a Gaussian distribution, therefore, it is fully characterized by its mean and variance, which we denote by $\kappa_1$ and $\kappa_2$. 
Both are functions of $N_{ch}$ and $b$.  
The crucial point is that the mean $\kappa_1$ is proportional to the multiplicity fluctuation $N_{ch}-\overline{N_{ch}(b)}$. 
It is proportional to the Pearson correlation coefficient $r_{N_{ch}}$ between $\delta p_t$ and $N_{ch}$ (straight blue line in Fig.~\ref{fig-1} right). 

The last step is to average over $b$. 
For this, one needs the probability distribution of $b$ at fixed $N_{ch}$, which is obtained using Bayes' theorem $P(b | N_{ch})p(N_{ch}) = P(N_{ch} | b) P(b)$ along the lines of Ref.~\cite{Das:2017ned}.
A simple calculation then shows that 
\begin{equation}\label{cumulants}
	\begin{aligned}
		\text{mean} \ \langle \delta p_t  \rangle &= \langle \kappa_1 \rangle_{b} \\
		\text{Var}(p_t) 
		&= (\langle \kappa_1^2 \rangle_{b} - \langle \kappa_1 \rangle_{b}^2)  + \langle \kappa_2 \rangle_{b} 
	\end{aligned}
\end{equation}
where angular brackets $\langle \dots \rangle_{b}$ denote the average over $b$ for fixed $N_{ch}$.
Two terms contribute to the variance: 
The first term originates from fluctuations of the impact parameter $b$ at fixed $N_{ch}$, and the second term is the true intrinsic (quantum) fluctuation, which is not the by-product of $b$ fluctuation. 

The expression in Eq.~(\ref{cumulants}) is finally fitted to ATLAS data using the three parameters : $\sigma_{\delta p_t}$, $r_{N_{ch}}$ and $\alpha$. Fig. \ref{fig-1} displays the model fit to the data (red lines) and the two separate contributions to Var($p_t | N_{ch}$). 
Our model precisely explains the steep decrease of the variance around the knee. 
Below the knee, both contributions have comparable magnitudes, but above the knee, the first term, due to impact parameter fluctuations,  gradually disappears, causing the steep fall in the variance data. 
This occurs because the effect of $b$ fluctuation becomes negligible in ultra-central region due to the strict lower limit of $b$ ($\geq 0$), where the distribution of $\delta p_t$ becomes a truncated Gaussian \cite{Das:2017ned,Samanta:2023kfk}. 

Fig. \ref{fig-2} displays the two-dimensional distribution of $[ p_t ]$ and $N_{ch}$ returned by our fit. The white curves represent 99\% confidence ellipses at fixed impact parameters and they are tilted with respect to the horizontal axis as in the hydrodynamic calculation in Fig. \ref{fig-1}. The tilts denote the positive correlation between $[p_t]$ and $N_{ch}$, characterized by $r_{N_{ch}}$. From this plot, it can be seen that the width of $[p_t]$ distribution is partly because several ellipses contribute at a given $N_{ch}$ (first term in Eq. \ref{cumulants}) and the other part comes from the vertical width of a single ellipse (second term in Eq. \ref{cumulants}). As a corollary, we also predict a slight increase of the mean transverse momentum $\langle \delta p_t \rangle $ with $N_{ch}$, denoted by the black line in Fig. \ref{fig-2}.

\section{Physical significance and thermodynamic interpretation}
\label{interpretation}

Our results imply that there is a strong correlation between $[p_t]$ and $N_{ch}$ at fixed impact parameter, which is quantified by the Pearson correlation coefficient $r_{N_{ch}}=0.676$. 
As explained in Sec.~\ref{simulation}, such a correlation is naturally present in a hydrodynamic description. 
Hence, the recent ATLAS data further support the common hypothesis that local thermalization is achieved in Pb+Pb collisions. 
It is the first such evidence which does not involve anisotropic flow. 

 Additionally, our analysis and methodology pave the way to separate the classical (geometrical) and quantum (intrinsic) fluctuation. Moreover, the slight increase of mean transverse momentum with multiplicity could be used to extract the speed of sound in QGP medium \cite{Gardim:2019brr}, which has recently been measured by the CMS collaboration with great precision \cite{CMS-PAS-HIN-23-003}. 

\section*{Acknowledgments}
R. S. is supported by the Polish National Science Center under grant NAWA PRELUDIUM BIS: PPN/STA/2021/1/00040/U/00001 and PRELUDIUM BIS: 2019/35/O/ST2/00357.
S.B and J.J are supported by DOE DE-SC0024602. M. L. thanks the S\~ao Paulo Research Foundation (FAPESP) for support under grants 2021/08465-9, 2018/24720-6, and 2017/05685-2, as well as the support of the Brazilian National Council for Scientific and Technological Development (CNPq).

\bibliography{ref}

\end{document}